\documentclass[12pt]{article}

\usepackage{graphicx,capt-of}
\usepackage{color}
\usepackage{bm}
\usepackage[font=Large]{subfig}
\usepackage{tabularx,ragged2e,booktabs,caption}
\usepackage{grffile}

\def\beq{\begin{equation}}
\def\eeq{\end{equation}}
\def\ba{\begin{eqnarray}}
\def\ea{\end{eqnarray}}

\begin{document}

\begin{center}
{\Large{\bf Finite temperature phase transition in the two-dimensional Coulomb glass at low disorders}} \\
\ \\
\ \\
by \\
Preeti Bhandari$^{1,2}$ and Vikas Malik$^3$ \\
$^1$Department of Physics, Jamia Millia Islamia, New Delhi 110025, India. \\
$^2$Department of Physical Sciences, Indian Institute of Science Education and Research, Mohali 140306, India. \\
$^3$Department of Physics and Material Science, Jaypee Institute of Information Technology, Uttar Pradesh 201309, India.
\end{center}

\begin{abstract}
We present numerical evidence using Monte Carlo simulations of finite temperature phase transition in two dimensional Coulomb Glass lattice model with random site energies at half-filling. For the disorder strengths ($W$) studied in this paper, we find the existence of charge-ordered phase (COP) below the critical temperature ($T_{c}(W)$). Also the probability distribution of staggered magnetization calculated at each W shows a two-peak structure at their respective critical temperature. Thus the phase transition from fluid to COP as a function of temperature is second order for all $W$. We find no evidence of a spin glass phase between a fluid and the COP. Further, we have used finite-size scaling analysis to calculate the critical exponents. The critical exponents at zero disorder are different from the one found at finite disorders, which indicates that the disorder is a relevant parameter here. The critical exponent for correlation length $\nu$ increases and $T_{c}$ decreases with increasing disorder. Similar behaviour for $\nu$ was seen in the work of Overlin et al for three dimensional Coulomb Glass model with a positional disorder. Our study also shows that other critical exponents are also a function of disorder.
\end{abstract}

\newpage

\section{Introduction}
\label{intro}

 Coulomb glass (CG) refers to disordered insulators where the electrons are localized and the interaction between these localized electrons is the long-range Coulomb interactions. This model is used to describe the properties of compensated doped semiconductors, ultrathin films and granular metals \cite{ref1,ref1a,ref1c}. Granular materials due to their tunable properties are an important class of thermoelectric materials. Recent studies on granular semiconductors have shown that they can be modelled by CG Hamiltonian \cite{ref1d,ref1e}. CG model study is divided into two parts depending upon the type of disorder present. In the first case, the sites form a regular lattice and there is random energy at each site (random energy model). In the second case, the sites are randomly positioned in the system so the model is referred to as the random-site model. In this paper, we concentrate only on the study of the random energy model. We first present the review of the work at high disorders. At very low temperatures, the Coulomb interactions remain unscreened which induces a soft gap (also called the Coulomb gap) in the single-particle density of states. Theoretical arguments \cite{ref2,ref2a,ref2b} predicts that the gap has a universal form $\rho(\varepsilon)\propto \vert \varepsilon-\mu \vert^{\delta}$ near the chemical potential $\mu$, where $\delta$ is the gap exponent ($\delta\geq D-1$ in D dimensions). There is strong numerical evidence \cite{ref3} to support the existence of the gap at $T \approx 0$. At finite temperature one expects that $\rho(\mu) \sim T^{\delta}$. This was seen numerically in two dimensions ($2D$) \cite{ref4,ref5} and three dimensions ($3D$) \cite{ref6}. Davies et al \cite{ref3}, in their paper, suggested that the presence of both disorder and long-range Coulomb interactions will make the system glassy. Although there have been experimental \cite{ref7a} and numerical \cite{ref7,ref8} evidence of glassy non-equilibrium effects on CG systems, it still remains unclear whether these effects are purely dynamical or it reflects a transition to an equilibrium glassy phase.
 
 In two-dimensional CG model, investigations have been done at higher disorders ($W>1$) to find the glass transition. In 1988, Xue and Lee \cite{ref23} performed Monte Carlo simulation on a 2D random site CG model, to investigate the glass transition by analyzing the single-site autocorrelation function and the equal time density-density correlation function as a function of momentum. They found similarity in the autocorrelation function with the 2D Ising spin glass model. They also found that the screening properties of the system cross over from the metallic screening at high temperature to dielectric screening at low temperatures, which is an indication of glassy behaviour. In contradiction to this, other numerical studies done by Menashe et al \cite{ref24} and Goethe et al \cite{ref24a} have shown the absence of any finite temperature glass transition in the two-dimensional CG. No finite-temperature thermodynamic glass transition has been observed experimentally as well.

 Recently, we have shown \cite{ref26} a first-order transition from COP to disordered phase at critical disorder ($W_{c}=0.2253$) for two-dimensional CG lattice model at zero temperature. The disorder is in the units of nearest neighbour electron-electron interaction which is taken to be unity (see Sec.\ref{MNS} for details). We will here study the effect of both temperature and disorder on two-dimensional CG lattice model to investigate the finite temperature phase transition. Study of RFIM using Monte Carlo simulation using single flip dynamics done using the Metropolis algorithm \cite{new} have shown that the equilibrium times becomes exponentially long as the temperature drops below the critical temperature. Since we are using the same Monte Carlo method, we have restricted ourselves to small system sizes ($L=16,20,24,32$). Our study at $T=0$ \cite{newa} showed that different disorder configurations had different $W_{c}$. The transition region for $L=16$ extended from $W=0.20$ to $W=0.50$. This has led us to restrict our calculations in this paper until $W\leq0.15$. For the studied disorders, we have obtained a second order transition from fluid to COP at a finite transition temperature which is a function of $W$. The transition temperature decreases as $W$ increases. We did not find any evidence of a spin glass phase between the fluid and the COP. To check the relevance of disorder on the phase transition, we have also studied the system at $W=0$. We find that the critical exponents at $W=0$ are nearly same as $2D$ Ising model as represented by Mobius et al \cite{ref13}. The critical exponent at $W \neq 0$ are different from $W=0$ which means that the disorder is relevant. This is similar to three-dimensional random field Ising model (RFIM) and Coulomb glass.   

 The paper is organized as follows. In Sec.\ref{LR}, we first review the relevant literature of 3D RFIM and CG. Then in Sec.\ref{MNS}, we discuss the Hamiltonian of the system and then our numerical simulation. The results obtained from the simulation and its interpretation are presented in Sec.\ref{reslt} and the conclusion in Sec.\ref{conclsn}.

 \section{Literature review}
\label{LR}
The random field Ising model is one of the well studied short range disordered models. The work on RFIM is divided into Gaussian and bimodal model depending on the way one chooses the random fields. At $T=0$, in both the models there exists a ferromagnetic phase at disorders less than the critical disorder ($W_{c}$) and a paramagnetic phase at higher disorders. The transition from the ferromagnetic phase to paramagnetic phase as a function of disorder and temperature is second order for the Gaussian model \cite{Aj,Ds,Tn,Jv}, but there are few arguments which predict the occurrence of first order transition as well \cite{Jm,Yong}. The experimental realization of RFIM is $3D$ dilute anti-ferromagnetic system in presence of magnetic field. For this system, Nowak and Usadel \cite{ref9} had shown numerically a phase diagram in which a spin glass phase exists between the COP and the paramagnetic phase for disorder strength $W < W_{c}$. But subsequent numerical studies and recent theoretical studies \cite{ref10} on RFIM have proved that no spin glass phase occurs at any disorder.

 A phase diagram on CG lattice model in $3D$ which is similar to one found in three-dimensional RFIM Gaussian model was given by Goethe et al \cite{ref6} and Barzegar et al \cite{Bar}. Both studies show that there exists a charge-ordered phase (COP) below a critical disorder strength ($W_{c}$). These studies also found a second order phase transition from fluid to COP with no spin glass phase in between. Moreover, they find three-dimensional CG to be in the same universality class as three-dimensional RFIM. For $W>W_{c}$, Barzegar et al have found a transition from plasma to glassy phase at very low temperatures. In contradiction to this Goethe et al and Surer et al \cite{ref11,ref12} have found no glassy phase for $W>W_{c}$. Mean field studies \cite{ref7,ref14,ref15,ref16} on three-dimensional CG system at high disorders predicts a replica symmetry broken equilibrium glass phase below a critical temperature in the presence of on-site disorder. Overlin et al \cite{ref17} studied random site model and showed that for all disorders (including $W=0$) phase transition as a function of temperature is second order. Their study shows that the COP changes to Coulomb glass with the increase in disorder at low temperatures. They found that the critical exponent $\nu$ increased with increasing disorder. This is important because if the CG system falls in the same universality class as RFIM, then the critical exponent values should be independent of disorder. 
 
 \section{Model and numerical method}
 \label{MNS}
  To study a CG, it is essential to have both disorder and the long-range Coulomb interactions in the system. The disorder in a system can be modelled in two ways. First, the local site energy is kept constant and disorder is brought into the picture through random positioning of the sites, which leads to varying interaction between them. This is known as the random site model. Another alternative is, one can choose to place the sites in a regular lattice and the disorder can be introduced through random local site energy, which is called the lattice model. This greatly reduces the calculation efforts as the distances are functions of the site index differences only. As there is a possibility that the two models give very different behaviour, so further study of both the models is definitely justified.
  
 We have used the classical $2D$ lattice model \cite{ref2} with the Hamiltonian
 \begin{equation}
 \label{eq1} H = \sum_{i} \phi_{i}S_{i} + \frac{1}{2} \sum_{i \neq j} J_{ij} S_{i} S_{j}.
 \end{equation}
 Here $\phi_{i}$ is the random on-site energy of site $i$ drawn from a uniform distribution on the interval $[-W/2,W/2]$, where $W$ is the disorder strength. $S_{i}=n_{i}-1/2$ is the pseudospin variable (with $n_{i} \in 0,1$) of site $i$. All the energies and temperature are measured in the units of $e^{2}/\kappa a$, where $\kappa$ is the dielectric constant and $e$ is the elementary charge. The distance is measured in the units of the lattice constant $a$. $J_{ij}=\frac{e^{2}}{\kappa \vert \overrightarrow{r_{i}} - \overrightarrow{r_{j}} \vert}$ is the Coulomb interaction between sites $i$ and $j$. We have used the periodic boundary condition so the distance between two sites is considered as the length of the shortest path between them. This leads to a cut off in the Coulomb interactions at the distance $L/2$ (where $L$ is the linear dimension of the system).
 
 To study the equilibrium properties of CG we have performed Monte Carlo simulation. A charge neutral system at half filling (on a square lattice of $N=L \times L$ sites) with completely random spin configuration $\lbrace S_{i}\rbrace$ was annealed down to $T=0.01$ using Metropolis algorithm \cite{ref27,ref28} and Kawasaki dynamics. Here a single Monte Carlo step (MCS) requires randomly choosing two sites of opposite spins for spin exchange. To calculate the change in energy ($\Delta E$) in the system, we have used the Hartree energy ($\varepsilon_{i}$) \cite{ref1,ref2} defined as the energy required to remove an electron from the occupied site or the energy gained by adding an electron to an empty site keeping the occupancy of other sites unchanged. It is given as
 \begin{equation}
 \label{eq2} \varepsilon_{i} = \phi_{i} + \sum_{j} J_{ij} S_{j}.
 \end{equation}
 The change in energy due to an electron hopping from site $i$ to site $j$ is then given by \cite{ref1}
 \begin{equation}
 \label{eq3} \Delta E = \varepsilon_{j} - \varepsilon_{i} - \frac{1}{r_{ij}}
 \end{equation}
 After each successful spin exchange, all the Hartree energies were updated accordingly. At high temperatures, we have done $3 \times 10^{5}$ number of MCS and at low temperatures, the number of MCS was increased up to $5 \times 10^{5}$ MCS.
 
 It is important to ensure that our results were obtained from the well-equilibrated system. So to find the equilibration time for our system, we calculated the time-dependent quantity called the spin-correlation function \cite{ref29,ref30}
 \begin{equation}
 \label{eq5} \chi_{SG}(t) = \frac{1}{N} \Bigg [\Bigg(\sum_{i} S_{i}(t_{eq}) S_{i}(2\hspace*{1mm} t_{eq})\Bigg)^{2} \Bigg]_{av},
 \end{equation}
 where $[...]_{av}$ is the ensemble average. $\chi_{SG}(t)$ in the limit of $t\rightarrow \infty$ reduces to the spin glass susceptibility \cite{ref29,ref30}
 \begin{equation}
 \label{eq6} \chi_{SG} =  \frac{1}{N} \Bigg[\sum_{ij} \langle S_{i} S_{j} \rangle^{2}\Bigg]_{av},
 \end{equation}
 where $\langle...\rangle$ is the thermal average. In eq.(\ref{eq5}), $t_{eq}$ is the equilibration time. If $t_{eq}$ is not sufficiently long then $\chi_{SG}(t)$ approaches $\chi_{SG}$ from above, behaving as a monotonically decreasing function. But if $t_{eq}$ is above the equilibration time, then $\chi_{SG}(t)$ tends to a constant and the fluctuations dies down. So to make sure that our runs were properly equilibrated we calculated $\chi_{SG}(t)$ at different simulation times which were divided into logarithmically spaced bins and $t_{eq}$ was considered that time where the three bins agree within the statistical errors. In Monte Carlo simulation the thermal average is replaced by the time average. All measurements were done after $t_{eq}$, at a constant time interval. 
 \section{Results}
 \label{reslt}
 \subsection{Order parameter}
 \label{op}
\begin{figure*}
	\includegraphics[width=4.5cm,height=4.5cm]{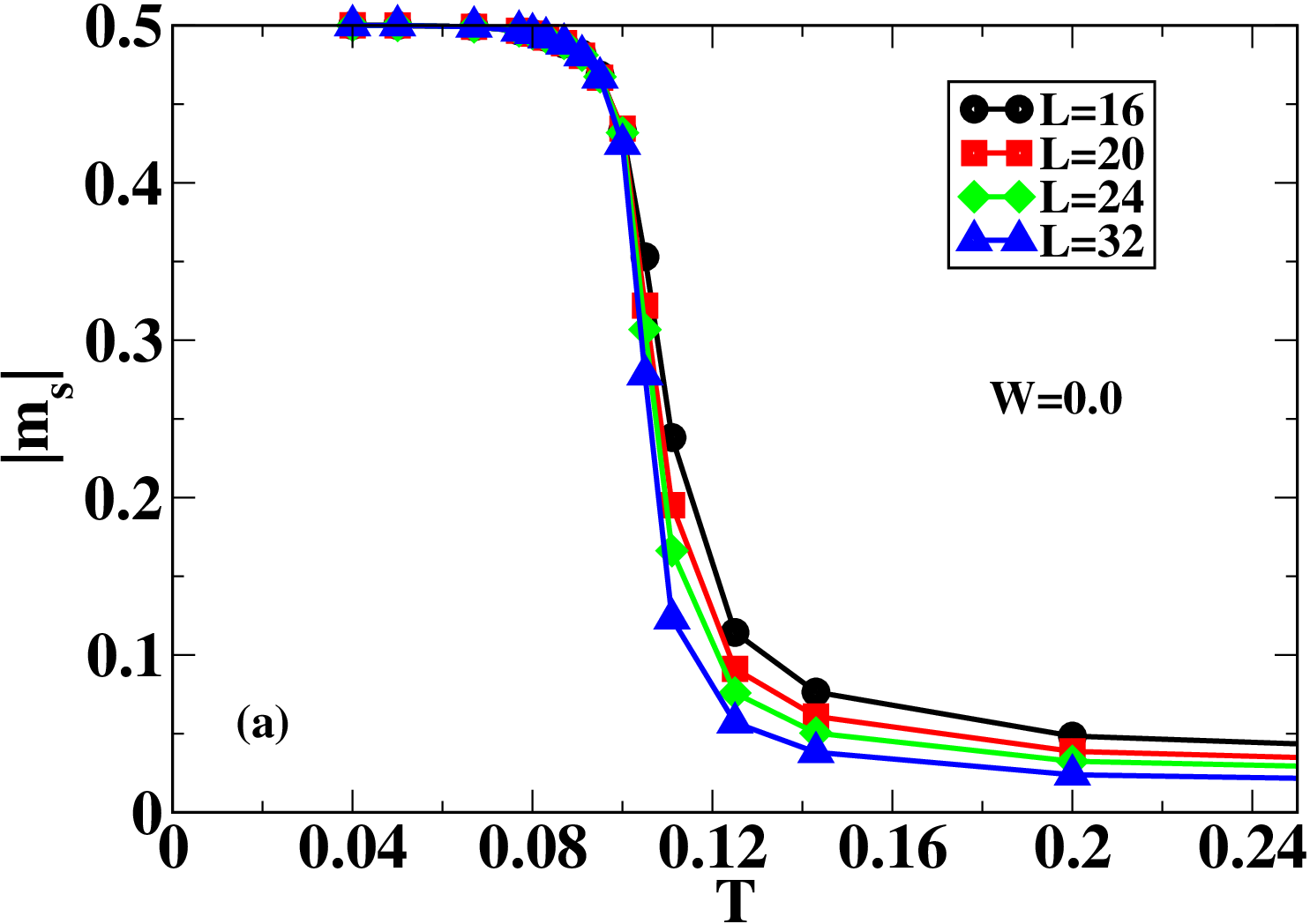}
	\includegraphics[width=4.5cm,height=4.5cm]{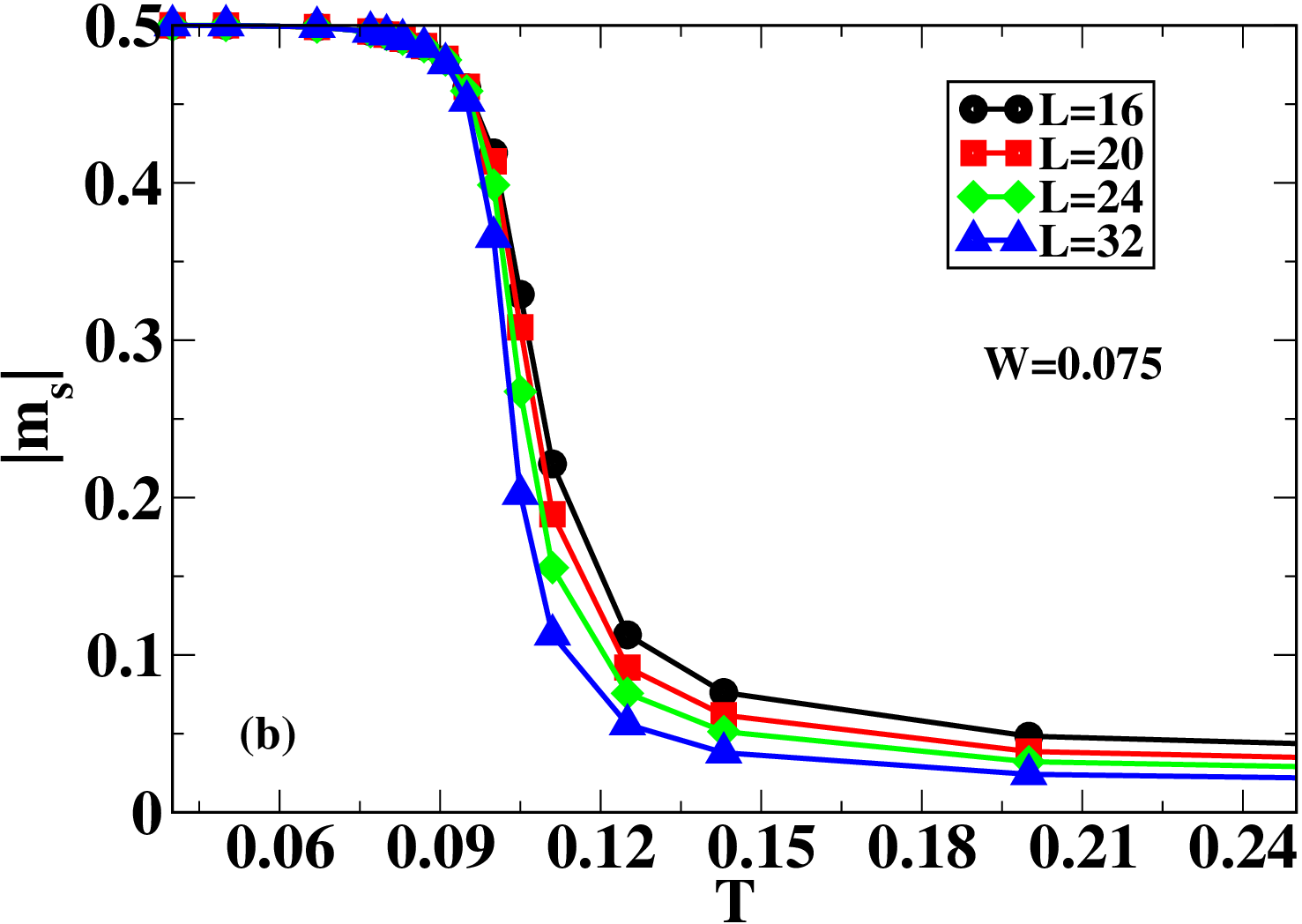}
	\includegraphics[width=4.5cm,height=4.5cm]{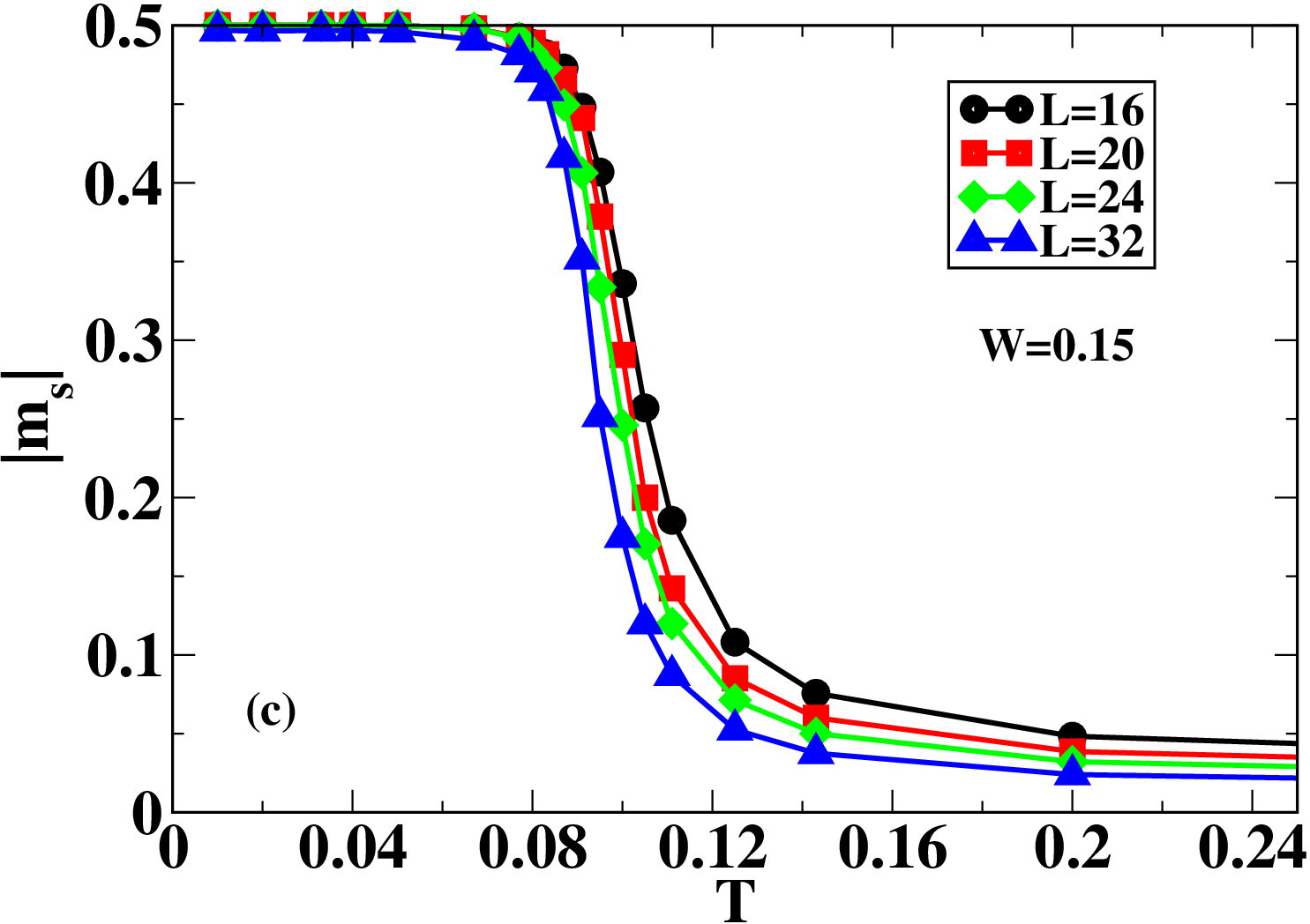}
	
	\includegraphics[width=4.5cm,height=4.5cm]{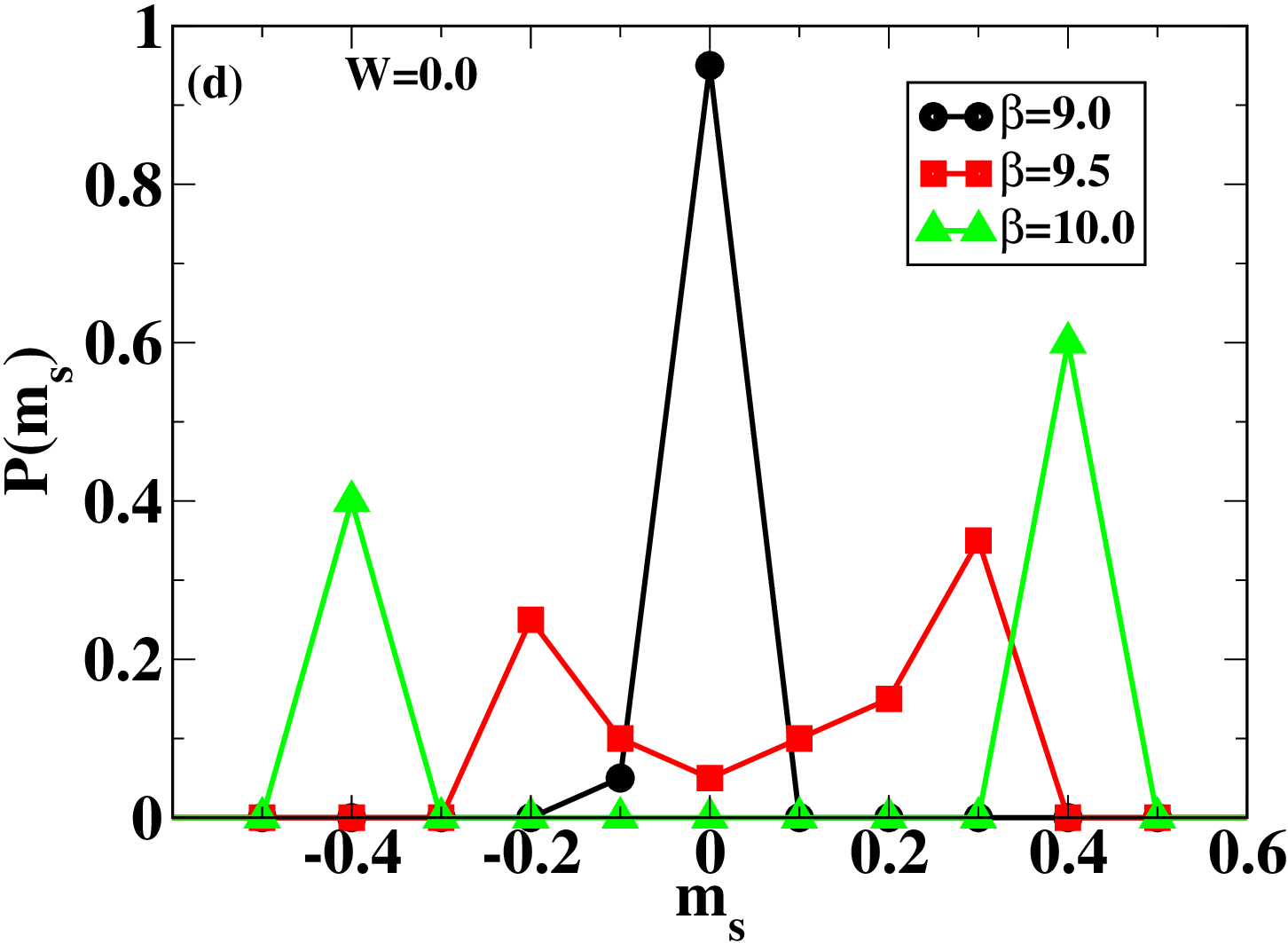}
	\includegraphics[width=4.5cm,height=4.5cm]{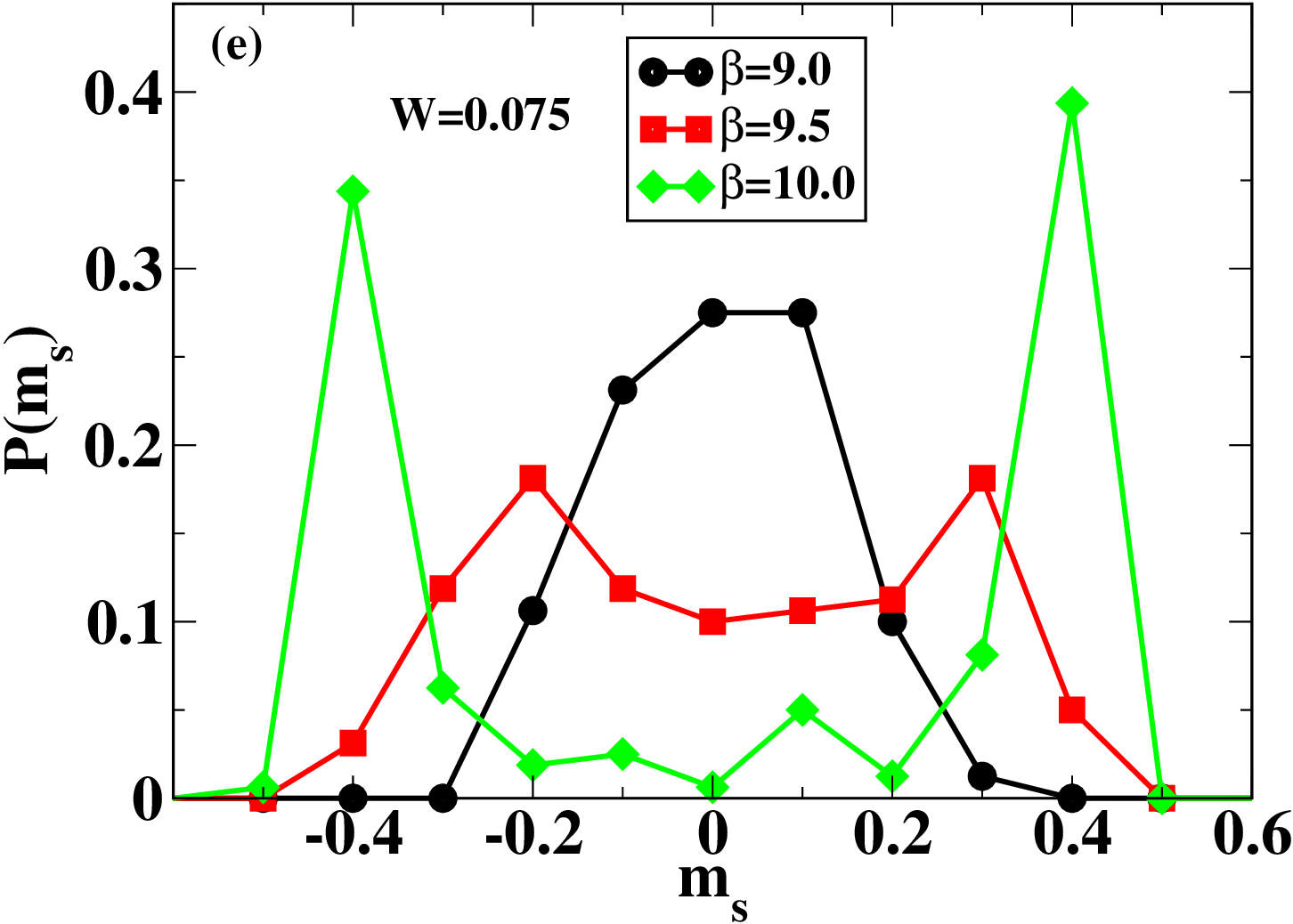}
	\includegraphics[width=5cm,height=5cm]{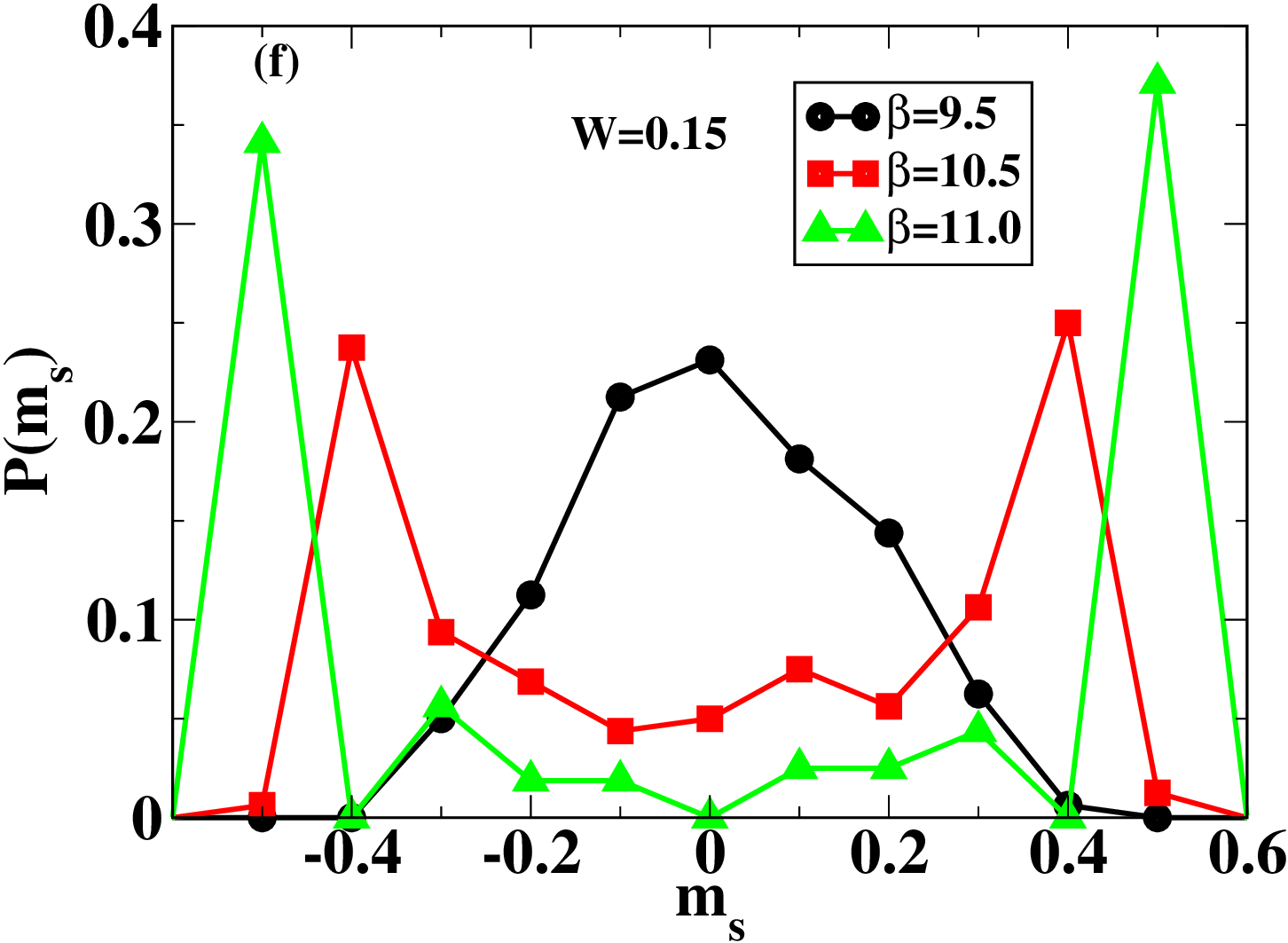}

	\caption{(Color online) (a)-(c) Behavior of $|m_{s}|$ as a function of temperature for different system sizes at each disorder. (d)-(e) The distribution of $P(m_{s})$ for $L=24$ at $W=0.0$, $W=0.075$ and $W=0.15$. No sign of coexisting phase found at any disorder strength.}
	\label{Fig1}       
\end{figure*}

  Our work is focussed on square lattice at half filling. For our system, in the absence of disorder, the ground state has two-fold degeneracy. This degeneracy is lifted by disorder and the COP is now unique. From our previous studies \cite{ref26}, we know that for $W<W_{c}$ at zero temperature, two-dimensional CG system exhibits antiferromagnetic ordering (COP). So one expects a transition from a high-temperature fluid phase to a low-temperature COP in this disorder regime. To investigate this transition at finite temperatures we use staggered magnetization, defined as
 \begin{equation}
 \label{eq8} m_{s} = \bigg[\bigg\langle\frac{1}{N}\sum^{N}_{i=1} \sigma_{i} \bigg \rangle\bigg]_{av}, 
 \end{equation}
 as an order parameter, where $\sigma_{i}=(-1)^{i} S_{i}$. In the paramagnetic phase, $m_{s}$ is zero as the spins fluctuate randomly, but as $T \rightarrow T_{c}$, spin ordering starts and $m_{s}$ becomes non-zero. The variation of $m_{s}$ at zero and finite disorders is shown in Fig.\ref{Fig1}. We have also calculated the distribution of staggered occupation $P(m_{s})$, to check the order of transition. For a first-order transition, one expects three peaks in $P(m_{s})$ at transition temperature ($T_{c}$) indicating phase coexistence. From Fig.\ref{Fig1} one can see that there is a single peak centred at $m_{s}=0$ when $T>T_{c}$ and at $T=T_{c}$ there are only two side peaks at $m_{s} \neq 0$ for $W=0.0$,$0.075$ and $0.15$, which indicates a second-order transition. 

\begin{figure*}
	\includegraphics[width=4.5cm,height=4.5cm]{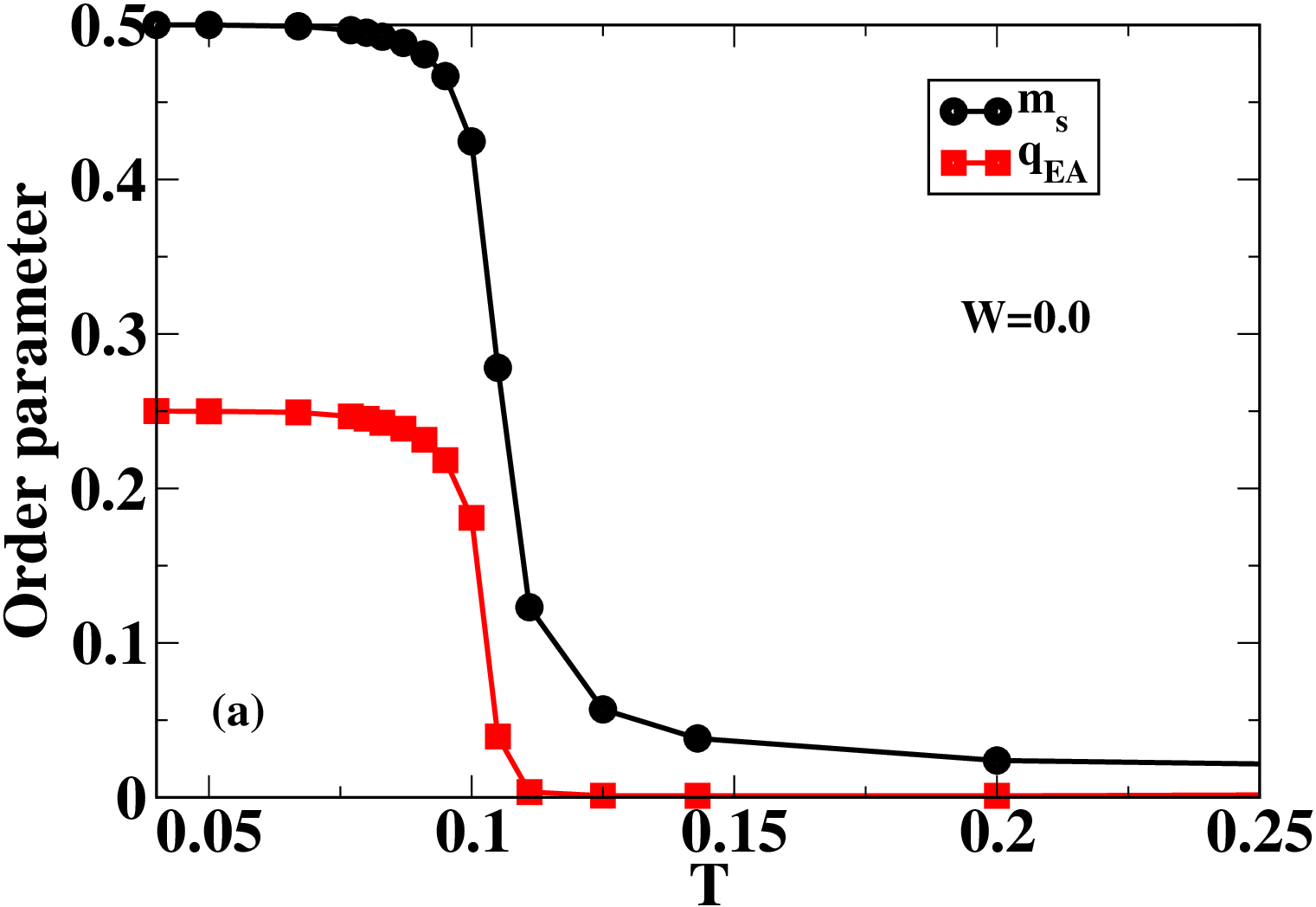}
	\includegraphics[width=4.5cm,height=4.5cm]{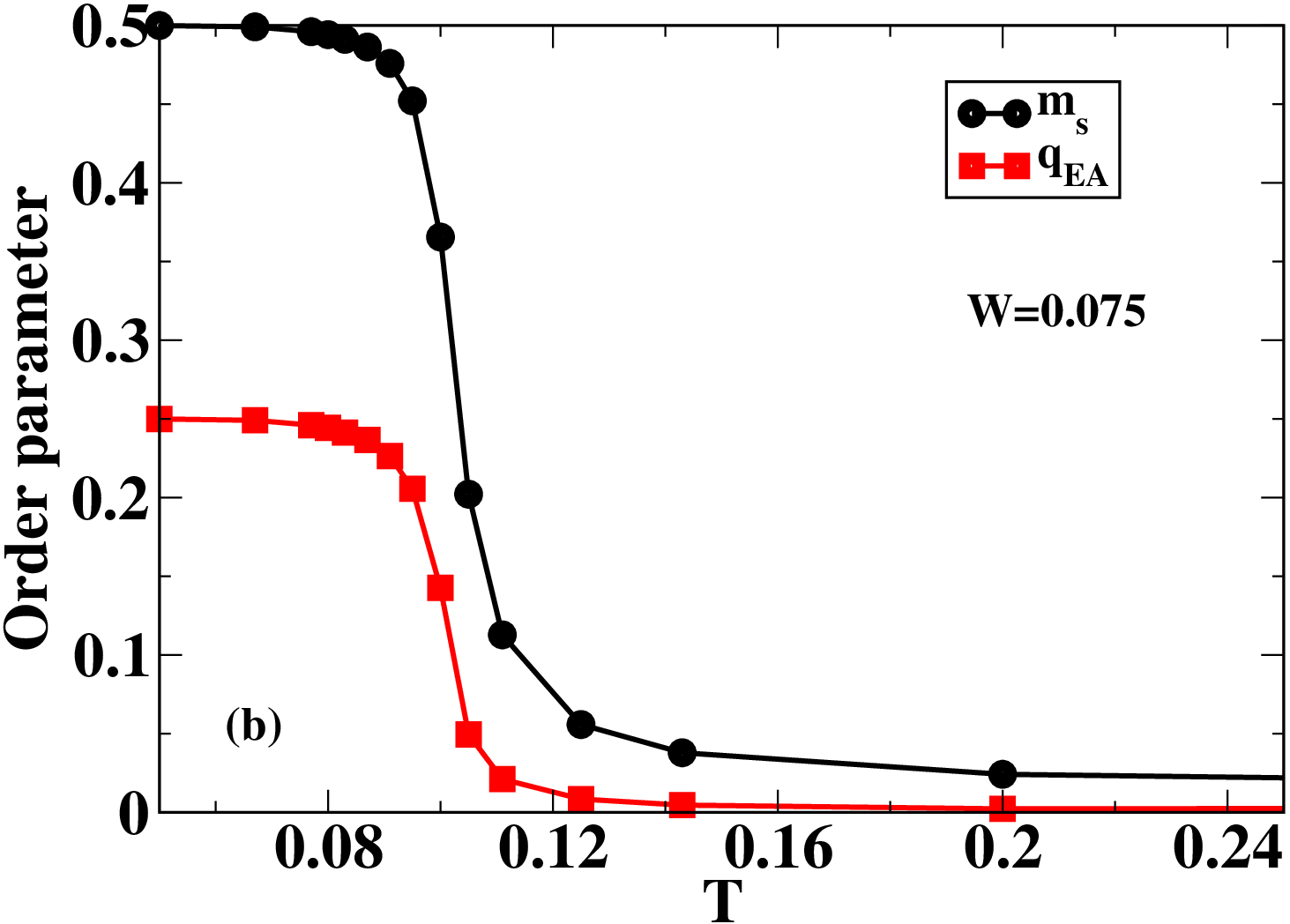}
	\includegraphics[width=4.5cm,height=4.5cm]{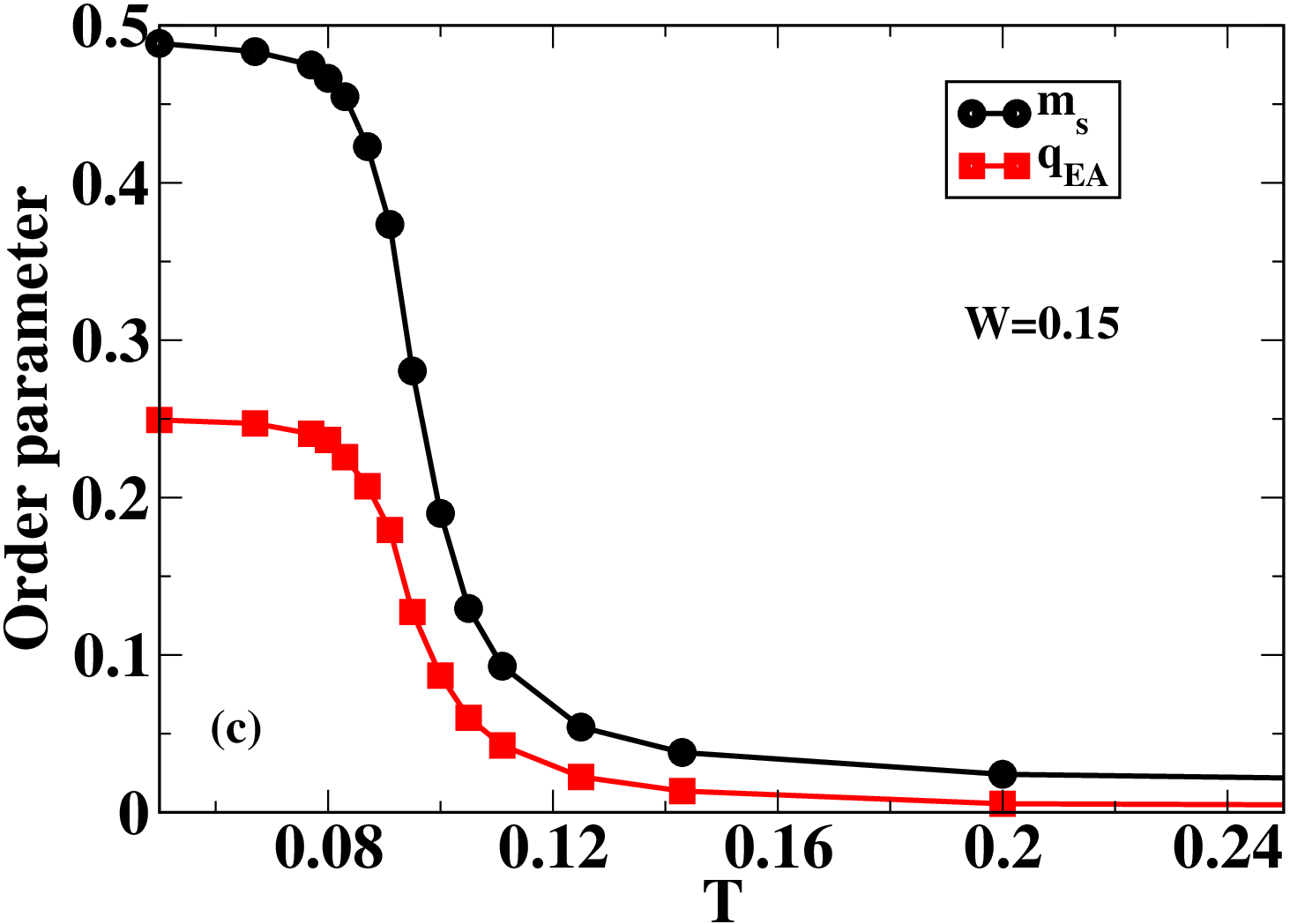}

	\includegraphics[width=4.5cm,height=4.5cm]{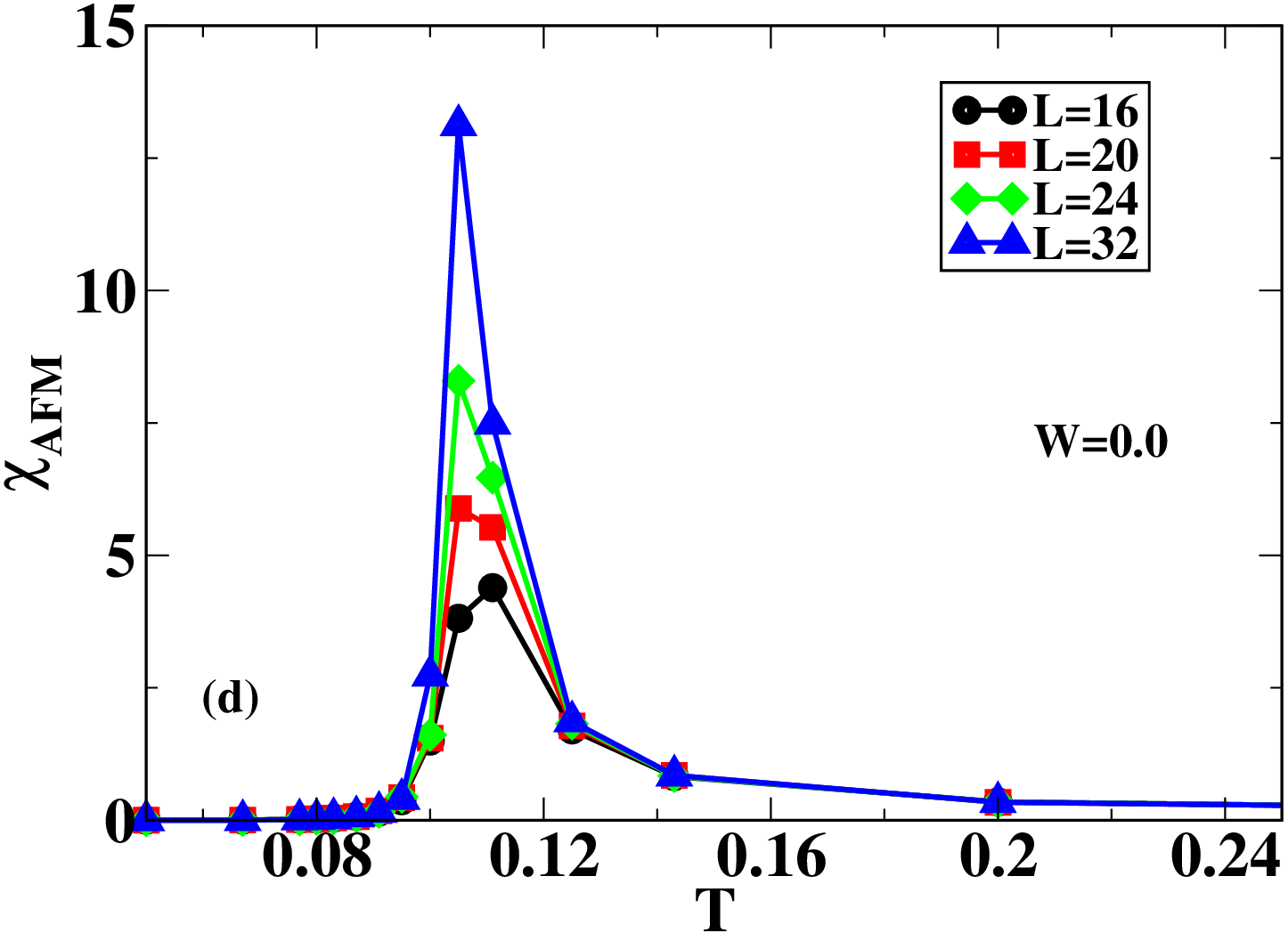}
	\includegraphics[width=4.5cm,height=4.5cm]{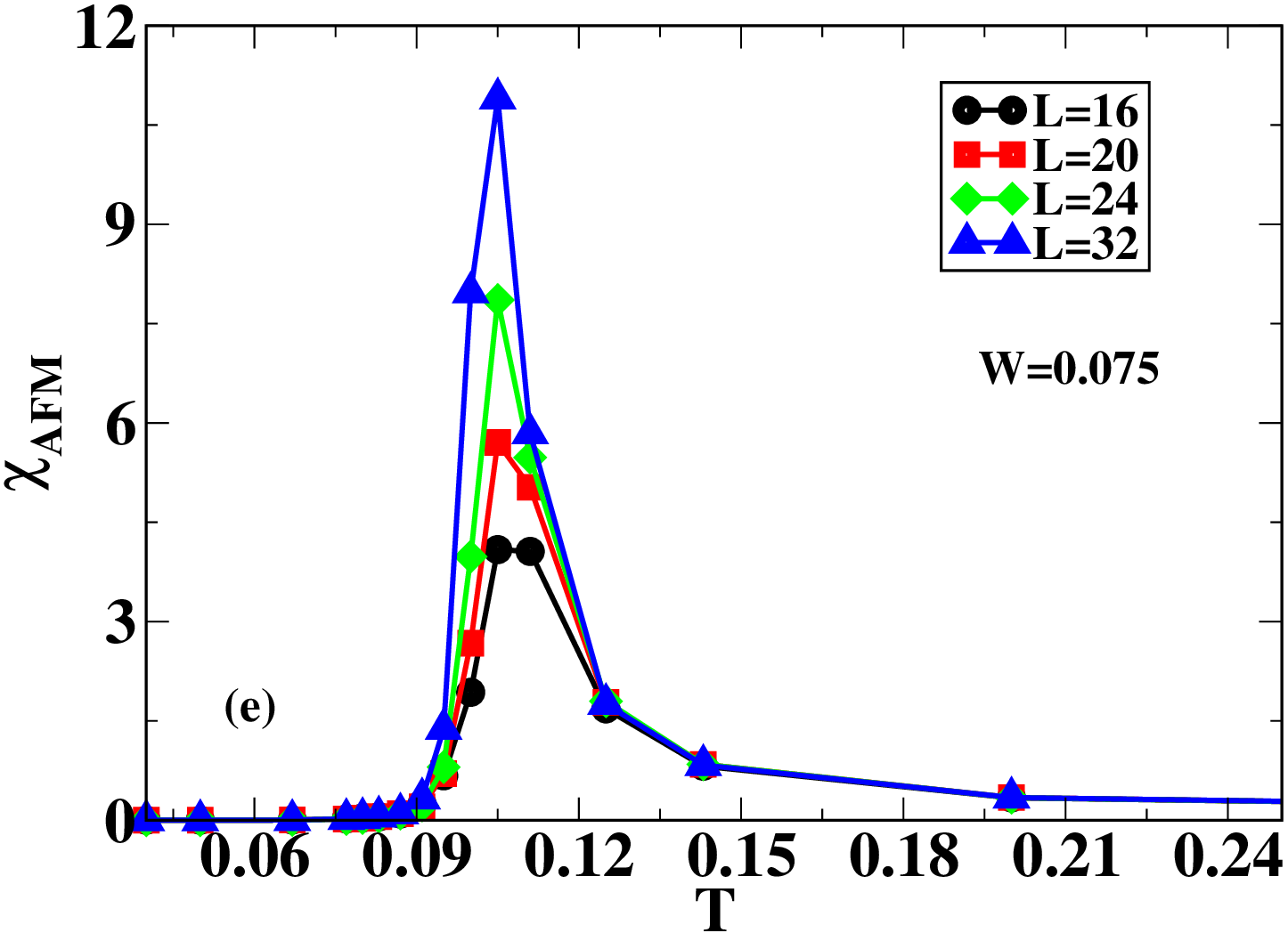}
	\includegraphics[width=5cm,height=5cm]{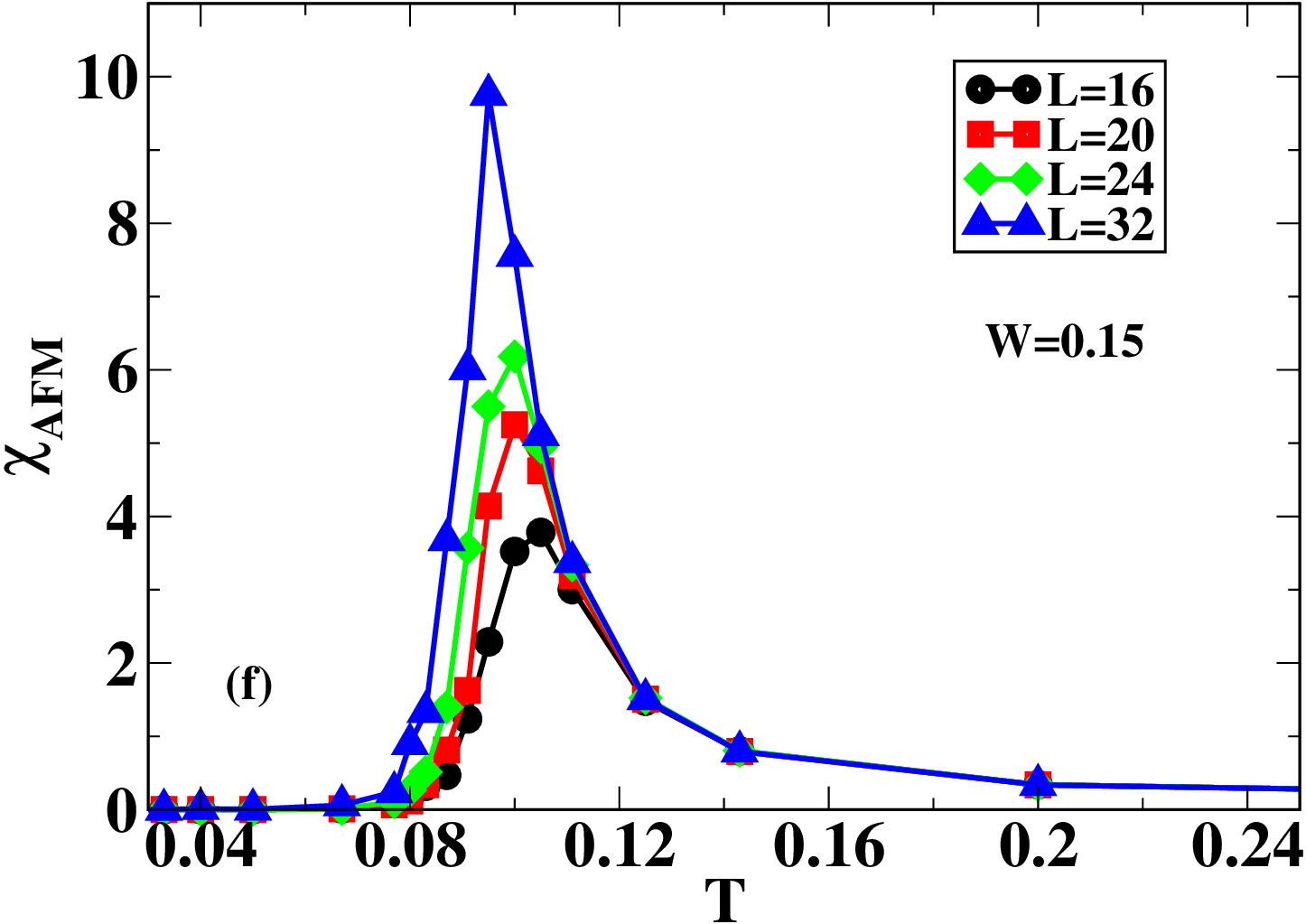}

	\caption{(Color online) (a)-(c) COP order parameter ($|m_{s}|$) and glass phase order parameter ($q_{EA}$) as a function of temperature are plotted at different disorder strengths  for $L=32$. (d)-(e) Temperature dependence of the antiferromagnetic susceptibility, $\chi_{AFM}(T)$ for all L at different disorders.}
	\label{Fig2}       
\end{figure*}

 In the study of spin-glasses, the parameter which has been studied most frequently is the one given by Edward and Anderson \cite{ref31} 
 \begin{equation}
 \label{eq9} q_{EA} = \Bigg[\frac{1}{N} \sum_{i=1}^{N}\langle S_{i} \rangle^{2} \Bigg]_{av}.
 \end{equation}
 The  Edward-Anderson order parameter $q_{EA}$ tells us about the tendency of a spin to have some preferential alignment. It provides information about the extent to which the spins or site occupation are frozen. So we have used $q_{EA}$ to measure glass like ordering in the system. $q_{EA}$ varies as $0 \leq q_{EA} \leq 0.25$ as $S_{i} = \pm 0.5$. At low temperature the spins are frozen, so the average orientation of spins will give the non-zero value of $q_{EA}$, but when the temperature is high, spins are fluctuating randomly and hence $q_{EA}$ vanishes. 
 If the system freezes to a glass phase (no long range ordering), then one expects $|m_{s}| \rightarrow 0$ and $q_{EA}  \rightarrow 0.25$. We have compared $q_{EA}$ and $|m_{s}|$ as a function of temperature at different disorders, to check the presence of glass phase. Fig.\ref{Fig2} shows the behavior of $q_{EA}$ and $|m_{s}|$ for $L=32$ at $W=0.0, 0.075$ and $0.15$. Both $q_{EA}$ and $|m_{s}|$ show sharp rise with temperature, at almost same $T_{c}$ indicating absence of glassy phase at these disorders.  
 
 \subsection{Susceptibility}
 \label{Xconn} 
 To confirm the occurrence of a phase transition at zero and small disorders, we have calculated antiferromagnetic susceptibility, defined \cite{ref13} as
 \begin{equation}
 \label{eq12} \chi_{AFM} = N [\langle m_{s}^{2} \rangle - \langle  m_{s} \rangle^{2}]_{av}
 \end{equation}
 In Fig.\ref{Fig2}, one can see the peak in $\chi_{AFM}(T)$ at $T_{c}$, which becomes sharper as the system size increases.
 
 \subsection{Critical Exponents}
 \label{fss} 
\begin{figure*}
\centering
	\includegraphics[width=3.5cm,height=3.5cm]{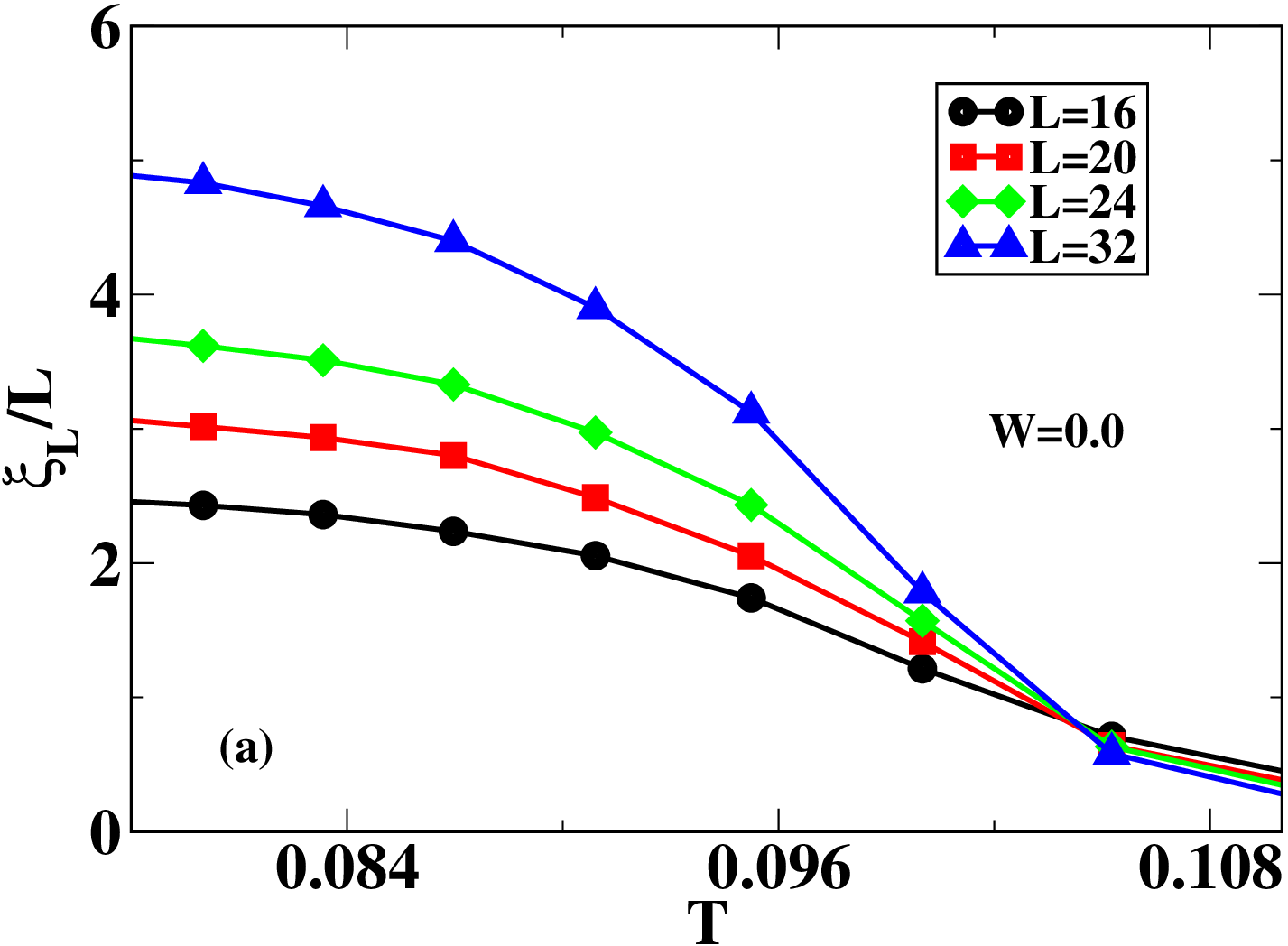}
	\includegraphics[width=3.5cm,height=3.5cm]{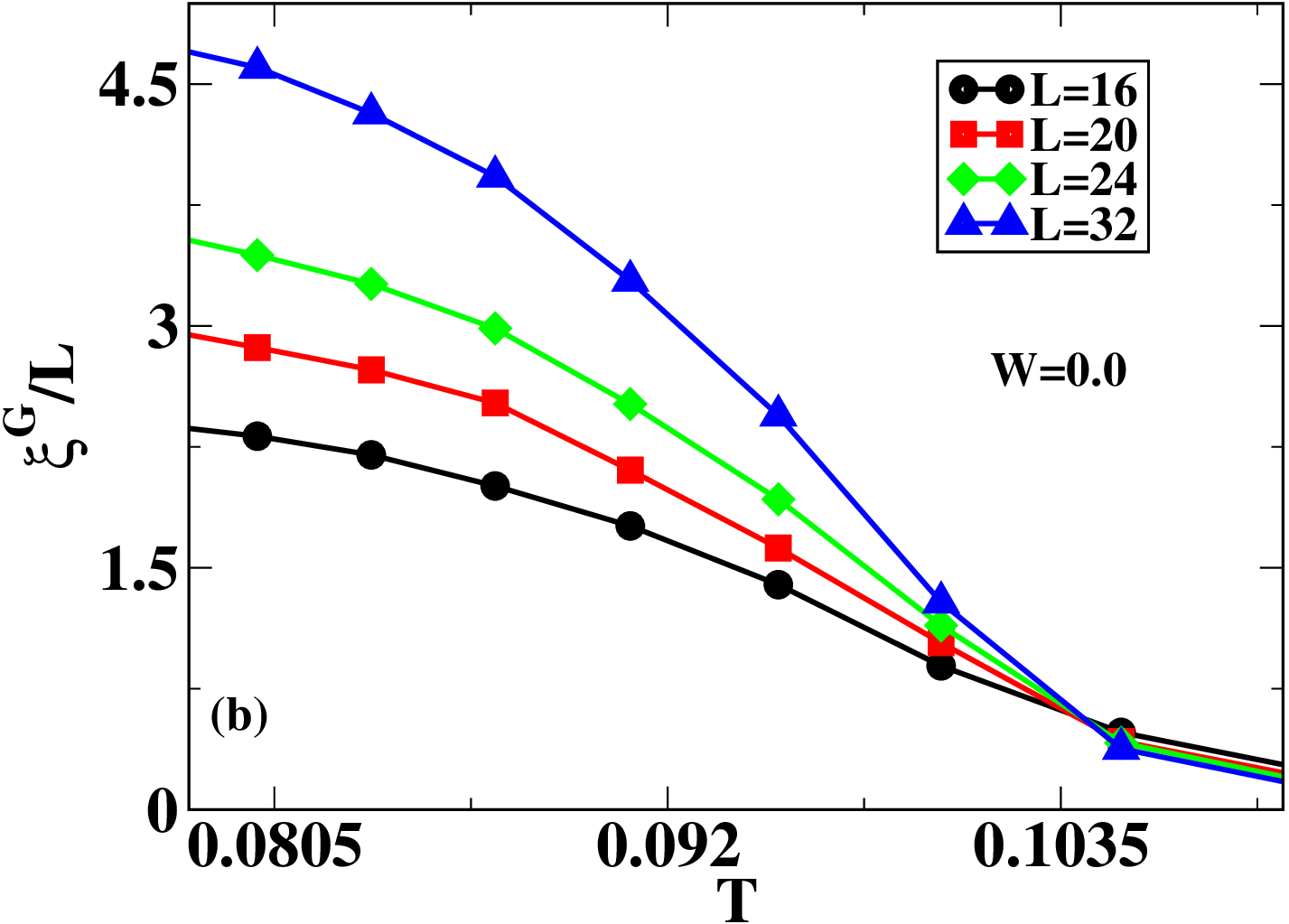}
	
	\includegraphics[width=3.5cm,height=3.5cm]{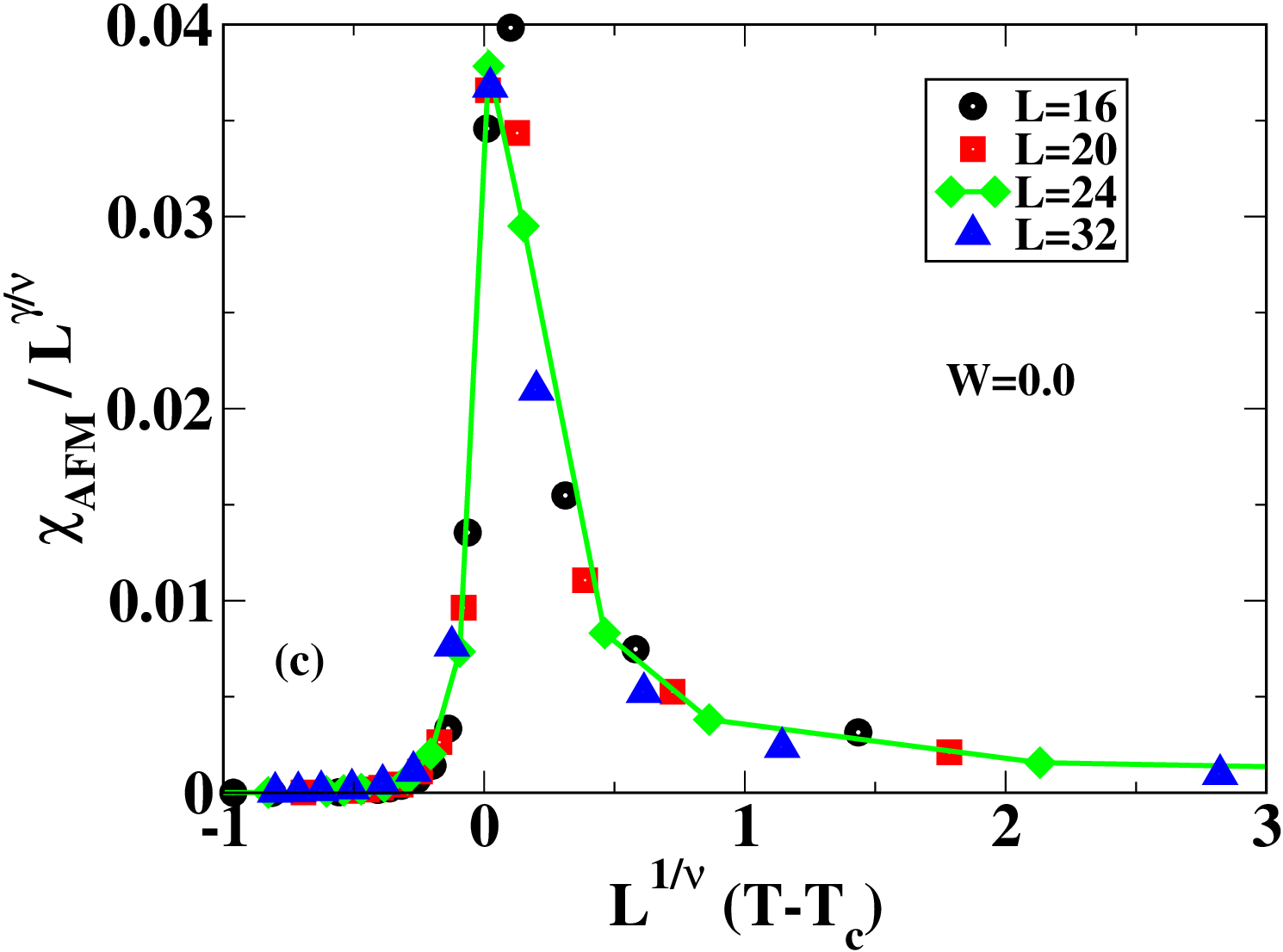}
	\includegraphics[width=3.5cm,height=3.5cm]{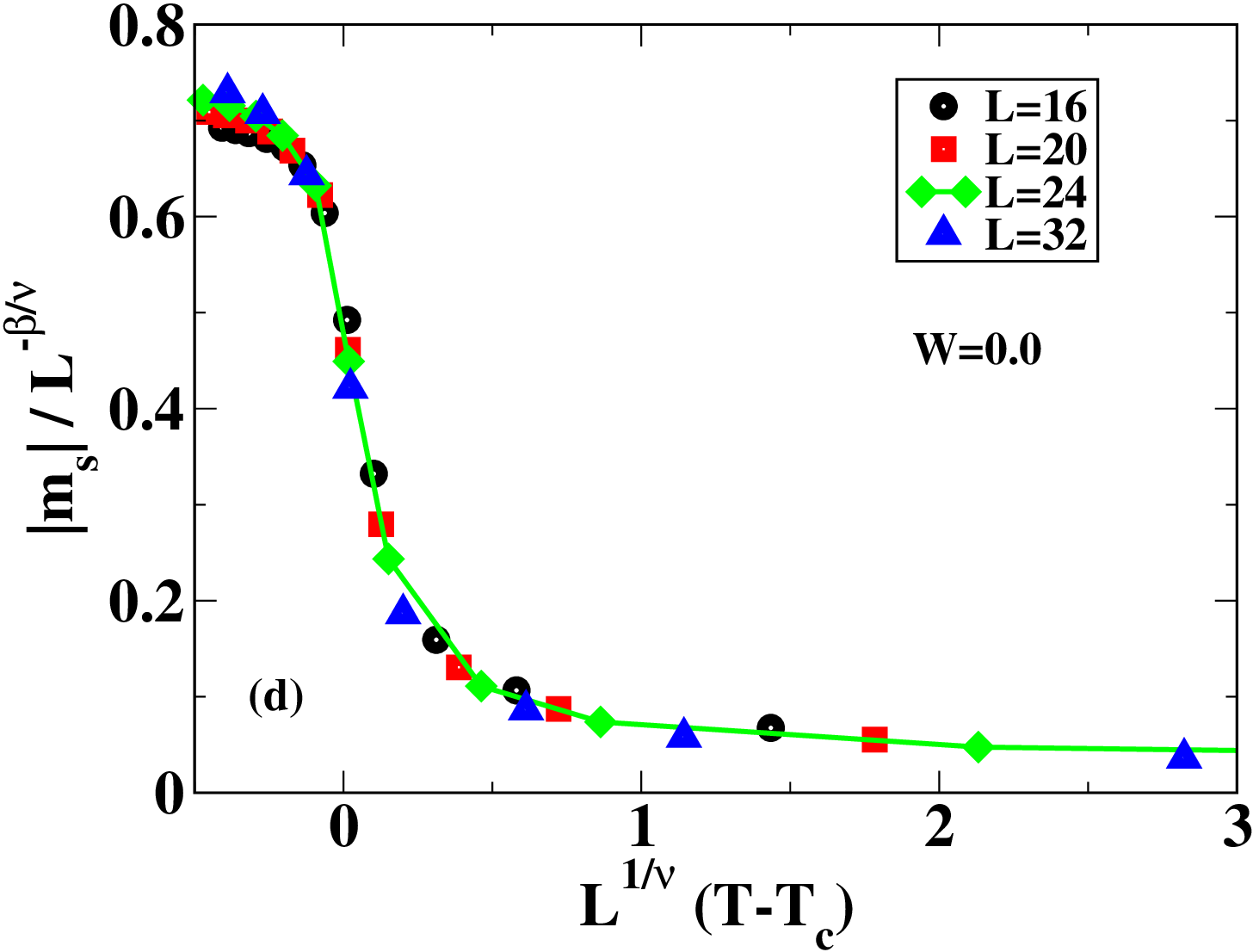}
		
	\includegraphics[width=3.5cm,height=3.5cm]{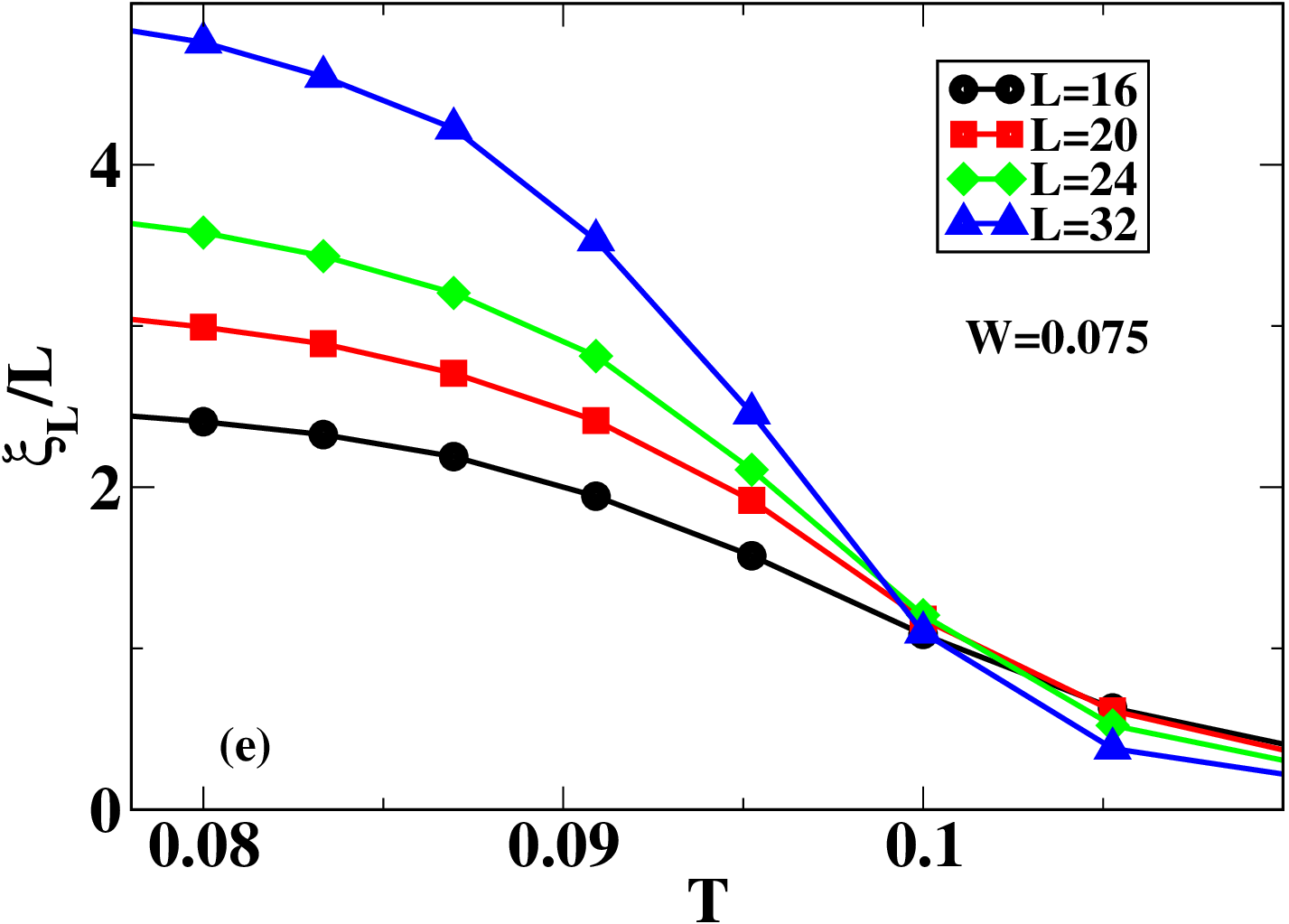}	
	\includegraphics[width=3.5cm,height=3.5cm]{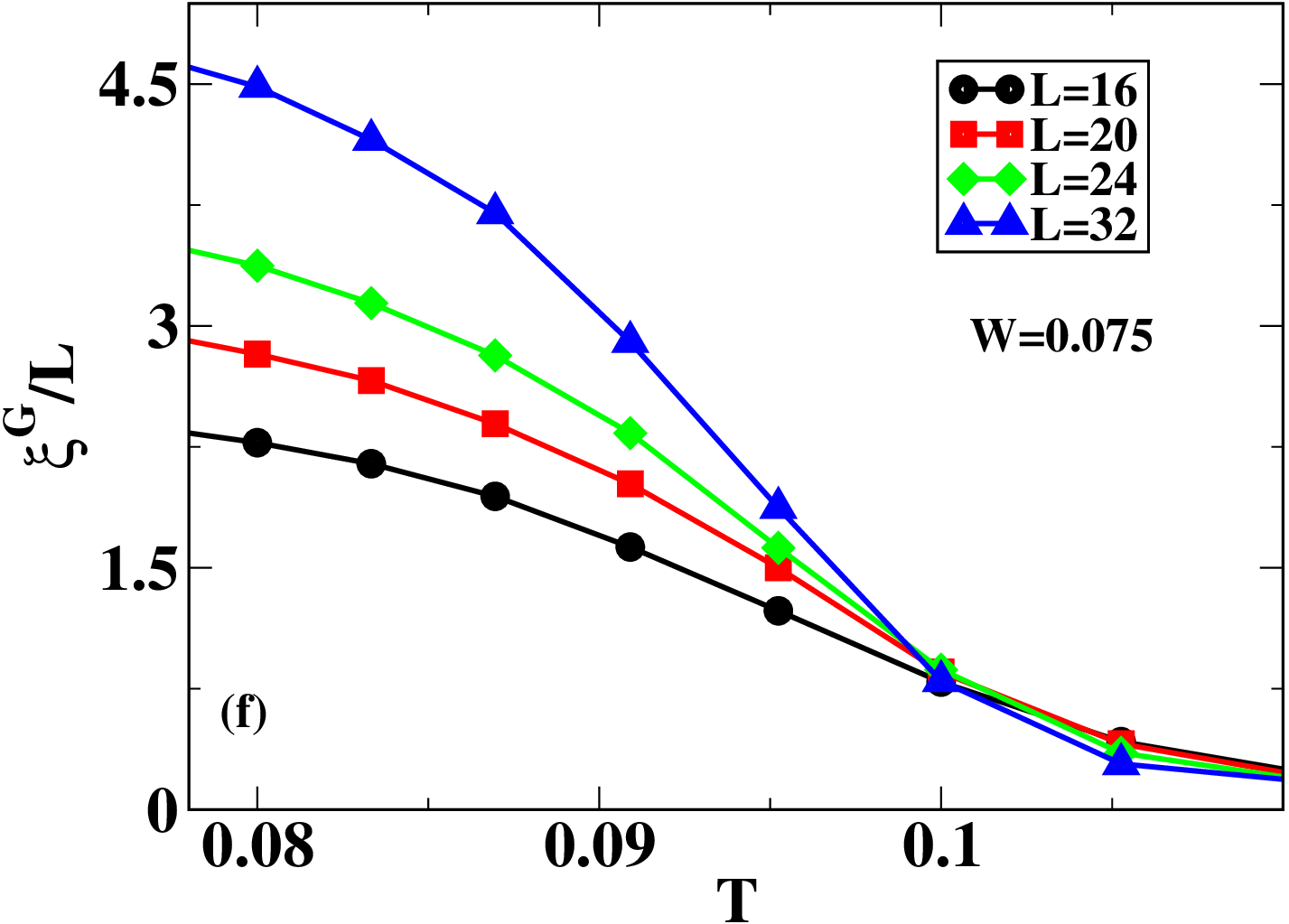}
	
	\includegraphics[width=3.5cm,height=3.5cm]{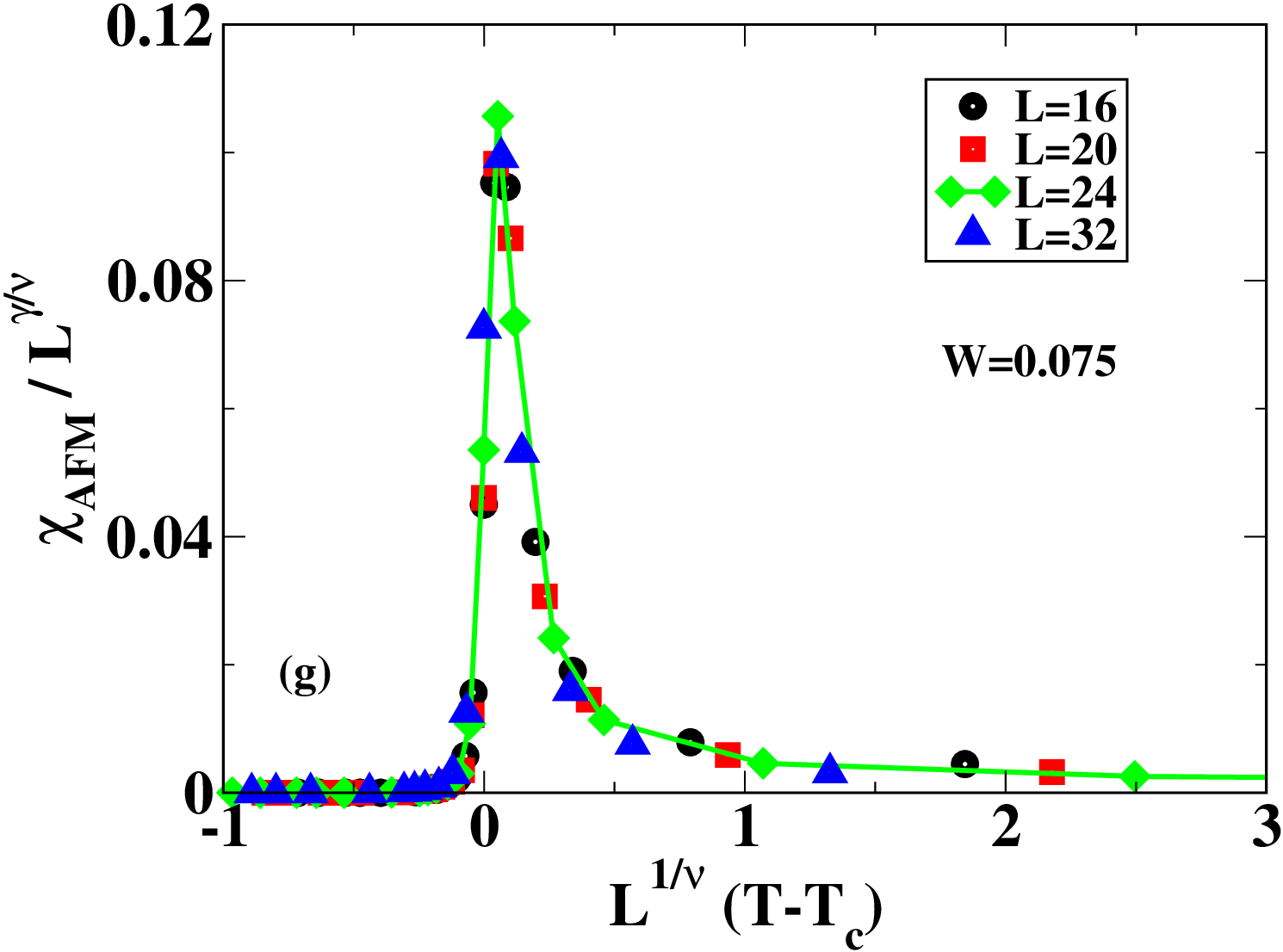}
	\includegraphics[width=3.5cm,height=3.5cm]{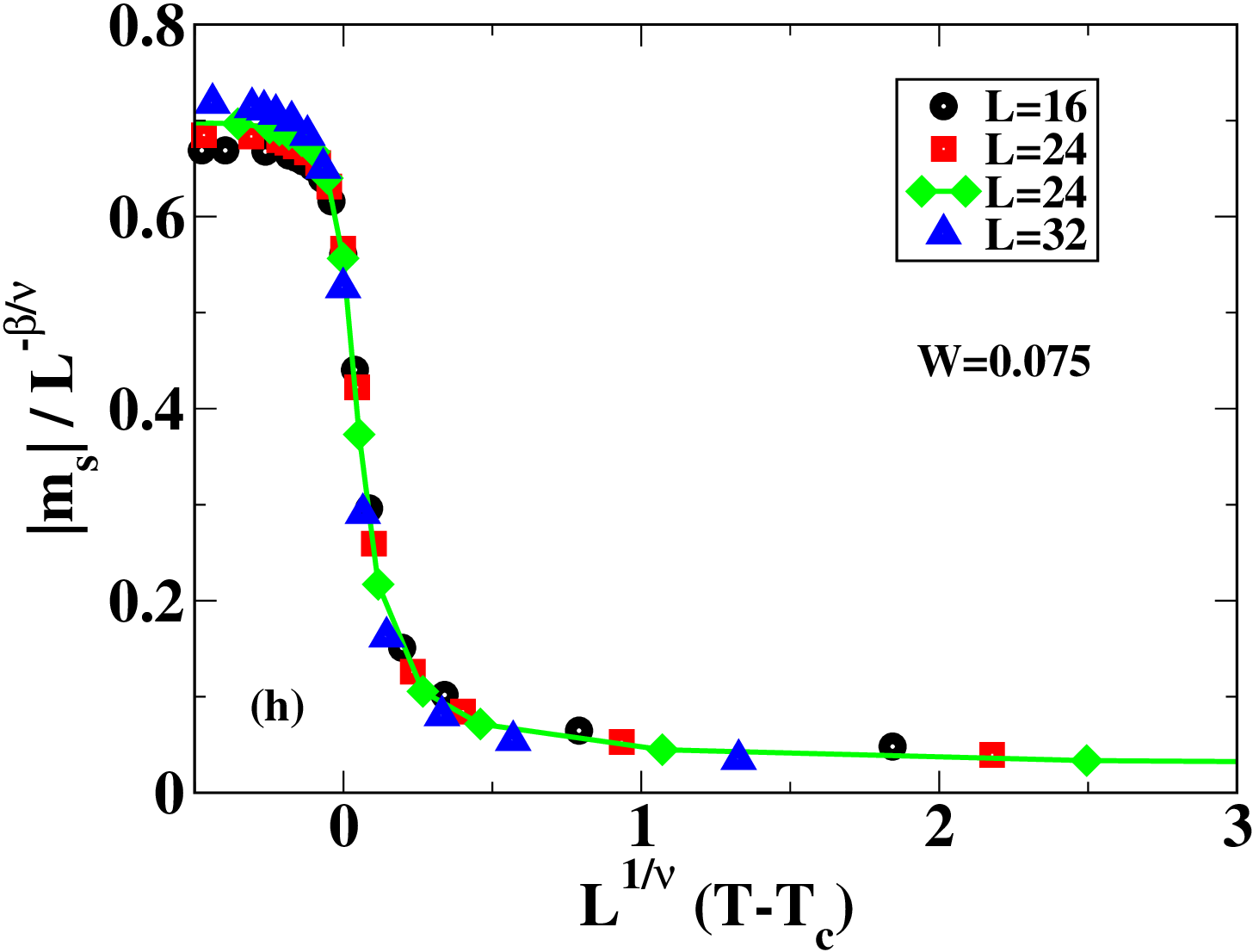}
	\caption{(Color online) (a)-(d) Finite size scaling analysis at $W=0.0$. (a) The correlation length in units of lattice size. (b) The spin glass correlation length in units of lattice size. (c) The scaling plot of the susceptibility $\chi_{AFM}$ with the parameters $T_{c}=0.104$, $\gamma/\nu=1.69$ and $\nu=1.024$. (d) The scaling plot of staggered magnetization $|m_{s}|$ with the parameter $\beta/\nu=0.120$. The full line in (c) and (d) connects only points for $L=24$ as a guide for the eyes. (e)-(h) Finite size scaling analysis at $W=0.075$. (e) The correlation length in units of lattice size. (f) The spin glass correlation length in units of lattice size. (g) The scaling plot of the susceptibility $\chi_{AFM}$ with the parameters $T_{c}=0.1001$, $\gamma/\nu=1.355$ and $\nu=1.34$. (h) The scaling plot of staggered magnetization $|m_{s}|$ with the parameter $\beta/\nu=0.105$. The scaling was done for $W=0.075$. The full line in (g) and (h) connects only points for $L=24$ as a guide for the eyes.}
	\label{Fig3}       
\end{figure*}

 To determine the transition temperature $T_{c}$ for charge ordered phase we measured the finite-size correlation length \cite{ref32,ref32a}
 \begin{equation}
 \label{tccal} \xi_{L} = \frac{1}{2sin(|k_{min}|/2)} \Bigg(\frac{\chi_{L}(0)}{\chi_{L}(k_{min})} - 1\Bigg)^{1/2}
 \end{equation}
 where $\chi_{L}(k) = N^{-1} \sum_{ij} \ [\langle\sigma_{i} \sigma_{j}\rangle]_{av} e^{-ik.r_{ij}}$ and $k_{min}=(\frac{2\pi}{L},0)$. By plotting $\xi_{L}/L$ for different $L$ as a function of $T$, one can obtain the critical temperature where the different $\xi_{L}/L$ plots cross. The crossing of our data of $\xi_{L}/L$ for different system sizes (see Fig.\ref{Fig3}) signals a phase transition at $T_{c}=0.1042$ for $W=0.0$ (which is close to the previous result $T_{c}=0.103$ \cite{ref13}), $T_{c}=0.1001$ for $W=0.075$, and $T_{c}=0.0888$ for $W=0.15$ (figure not shown). As expected the transition temperature decreases with an increase in disorder.
 
  We have also checked this crossing by replacing $\langle \sigma_{i} \sigma_{j} \rangle$ in eq.(\ref{tccal}) with the "spin-glass" correlation function
 \begin{equation}
 G(r_{ij}) = [\langle S_{i}S_{j} \rangle ^{2}]_{av}
 \end{equation}
 as suggested by Ballesteros et al \cite{ref32}. The crossing temperatures obtained from both the methods differ by less than $1\%$. This indicates that there is no spin glass phase between COP and fluid phase in the studied disorder strengths for two-dimensional CG. Similar conclusions were drawn in three-dimensional CG work of Goethe et al \cite{ref6}.
 
 Next, we analyze the finite size scaling behaviour of $\chi_{AFM}$ near the critical temperature $T_{c}$. From the standard finite size scaling relation we know \cite{ref33}
 \begin{equation}
 \label{Xeq} \chi_{AFM} (T,L) \approx L^{\gamma/\nu} \hspace*{1mm} \tilde{\chi} \hspace*{1mm} [L^{1/\nu} (T-T_{c})]
 \end{equation}
 The exponent $\gamma/\nu$ is extracted by plotting a least-squares straight-line fit of $\chi^{*}_{AFM}$ versus $L$ in the log-log plot where $\chi^{*}_{AFM}$ is the peak value of $\chi_{AFM}$ at the respective system sizes. The value of $\gamma/\nu$ at different disorders is summarized in Table \ref{T1}. With the estimate of $T_{c}$ and $\gamma/\nu$, $\chi_{AFM}$ was scaled using eq.(\ref{Xeq}) as shown in Fig.~\ref{Fig3} for $W=0.0$, $W=0.075$ and $W=0.15$ (figure not shown). We could not calculate the correlation length exponent $\nu$ directly so the value which provided the best fit was used. We have also scaled the data of staggered magnetization at different disorders using the scaling relation \cite{ref33}
 \begin{equation}
 \label{mseq} m_{s} (T,L) \approx L^{-\beta/\nu} \hspace*{1mm} \tilde{m} \hspace*{1mm} [L^{1/\nu} (T-T_{c})]
 \end{equation}
 and the scaled data for $W=0.0$, $W=0.075$ are shown in Fig.~\ref{Fig3} ($W=0.15$ not shown). The value of the critical exponents at different disorders is summarized in Table \ref{T1}. Numerical simulation at $W=0$ for finite temperature in two-dimensional CG system \cite{ref13} has claimed to obtain the values of the critical exponent consistent with those of the Ising model with short-range interactions. Our estimated values of critical exponents at $W=0$, are close to the $2D$ Ising model exponents as well.
 %
 \begin{table}
 	\caption{Critical exponents for the 2D Coulomb glass.}
 	\label{T1}       
 	\center
 	\begin{tabular}{lllll}
 		\hline\noalign{\smallskip}
 		$W$ & $T_{c}$ & $\gamma/\nu$ & $\nu$ & $\beta/\nu$ \\
 		\noalign{\smallskip}\hline\noalign{\smallskip}
 		0.0 & 0.1042 & 1.69 & 1.024 & 0.120 \\ 
 		0.075 & 0.1001 & 1.355 & 1.34 & 0.105\\ 
 		0.15 & 0.088 & 1.338 & 1.37 & 0.07\\
 		\noalign{\smallskip}\hline
 	\end{tabular}
 \end{table}
\section{Conclusions}
\label{conclsn}
 In continuation of our earlier work \cite{ref26} on two-dimensional CG model at $T=0$, we have here studied this model at finite temperature and small disorder strength. Previously we found a first-order transition from COP to disordered phase with the increasing disorder at zero temperature. Here we present evidence of finite temperature second order phase transition from fluid to COP at low disorders ($W \leq 0.15$). At zero disorder, our critical exponents match with the study done by Mobius et al \cite{ref13}. We find no evidence of spin glass phase between the COP and fluid phase for the studied disorders as claimed in three-dimensional CG system \cite{ref6} as well. An important observation in the current work is that the critical exponents at $W=0$ are different from $W \neq 0$ which suggests that for the finite temperature phase transition disorder is a relevant variable. Moreover, we found that $\nu$ is increasing with increasing disorder, which was also seen in the work of Overlin et al  \cite{ref17} for three-dimensional CG model with the positional disorder. The value of critical exponent $\beta$ tends to zero as the disorder increases. At zero temperature we found \cite{ref26} $\beta=0$ at the critical disorder $W_{c}=0.2253$. The ratio $\gamma/\nu$ is also decreasing in our case. This shows that the critical exponents are a function of the disorder. This behaviour is completely different from the RFIM case where the critical exponents are the same for $0<W<W_{c}$.
\newpage


\newpage


\end{document}